# Participatory Budget Allocation Method for Approval Ballots

[1]Rutvik Page, [2]Arnav Doifode, [3,*]Jitendra Tembhurne, [4]Aishwarya Sagar Anand Ukey

[1]University of Chicago Booth School of Business, Chicago, Illinois, United States 60611

[2]New York University, Brooklyn, New York, United States 10012

[3,4]Department of Computer Science and Engineering, Indian Institute of Information Technology, Nagpur, Maharashtra, India 441108

[1]rutvikpage1999@gmail.com, [2]arnav.live@gmail.com, [3]jtembhurne@iiitn.ac.in, [4]aukey@iiitn.ac.in

[*]**Corresponding Author:** jtembhurne@iiitn.ac.in

## Abstract

In this paper, we study the problem of Participatory Budgeting (PB) with approval ballots, inspired by Multi-Winner Voting schemes. We present generalized preference aggregation methods for participatory budgeting, especially for finding seemingly fair budget allocations. To achieve this, we generalize such preference aggregation methods from the well-known methods, namely the Sequential Chamberlin Courant rule and the Sequential Monroe Rule in the realm of social choice theory. Further, we provide an experimental evaluation of the preference aggregation methods using an impartial culture method of preference generation and study the extent to which such polynomial time algorithms satisfy one of the most popular notions of fairness called proportional representation.

## 1. Introduction

Aggregation of preferences to ensure maximum voter satisfaction is a central problem in social choice theory. The inherent problem encountered in this pursuit is the existence of conflict between voters' opinions. Participatory budgeting addresses the problem of allocation of funds to projects of varying costs while spending not more than some budget limit by aggregating voter preferences fairly. In recent years, participatory budgeting has become an important exercise for public budgeting decisions and has been utilized in several cities across continents such as Paris, USA, etc. Most commonly, the exercise is carried out on the following lines - Municipal Corporation elicits a group of projects that it finds fit to fund and requests the voters of the particular jurisdiction to put forth a group of projects that they would like to fund. The Municipal Corporation then runs some choice aggregating algorithm to find such a subset of projects that somehow seem "fair" and also satisfy the budget limit. There has been a surfeit of effort to recognize fair budget allocation using axioms and to design algorithms that correspondingly output such budgets that satisfy these axioms.

The fundamental problem addressed by participatory budgeting is the division of a given fixed budget to fund a subset of projects based on voter preferences. Essentially, the goal is to select a subset of items from a set of items, each with a set cost to maximize voter satisfaction. A simpler instance of this problem is the selection of a committee with $k$ seats: in this case, the budget is $k$ and every candidate costs exactly *one* unit. The main avenues that need to be answered while considering participatory budgeting as a way to exercise fair allocation of resources are - 1) defining what exactly is *fair* and 2) how to design an algorithm that outputs a *fair* budget. One of the widely studied criteria for fairness in preference aggregation is Proportional Representation (PR). The Proportionally Representative Budget (PRB) ensures that a set of voters that is both cohesive in its opinion and is large enough to deserve representation in budgets is not overlooked by the preference aggregating algorithms.

The fair division of budgets is a central problem that virtually every institution has to address. This is especially true in cases where democratic institutions are expected to work efficiently. In such cases, one way to arrive at a budget that maximizes social satisfaction is to design a budget by aggregating the preferences of individual voters. Participatory budgeting addresses this exact conundrum by laying out the exact parameters of fairness and then aggregating the opinions of voters in the form of votes to arrive at an efficient and acceptable consensus over budgets.

As mentioned, the problem of designing *fair budgets* is to define the exact meaning of *fairness* and then design algorithms that output such efficient budgets. An encumbrance that is well known in solving such problems is that of NP-Hardness. Essentially, if we set parameters for fairness to be stringent, it turns out that the algorithms that find these budgets have a very large running time; that is, the problem of finding efficient budgets becomes NP-Hard. As a workaround, we adopt some well-known polynomial time preference aggregation algorithms to find budgets while sacrificing the strictness of *fairness*.

To the best of our knowledge, no attempt exists to study the empirical behavior of polynomial time preference aggregation methods to select a proportionally representative budget. In effect, our contributions can be summarized as follows:

- Definition of one new axiom - a fairness criterion.
- Adaptation of the Sequential Chamberlin-Courant Rule and Sequential Monroe Rule for the election of a budget constrained by a fixed budget.
- The empirical study of the proposed algorithm for electing budgets adherent to the proposed fairness axioms.
- Inference using visual aids to prove the efficacy of Standard Transferrable Voting to elect a proportionally representative budget.

In the following sections, we present the related works in Section 2 to show the current progress in the field of participatory budgeting, both from a theoretical and practical application standpoint. In Section 3, we proposed the notations, preliminaries, definitions, and framework of the proposed system. We present our inferences from experiments in Section 4. Finally, we conclude the paper.

## 2. Related Works

The origin of participatory budgeting goes back to 1989. The first country to implement participatory budgeting in the system was Brazil. The adoption of participatory budgeting in Brazil led to revolutionary developments [1-3], which inspired research and analysis of the system. Participatory budgeting is also used in some local government units in the State of Kerala, India to increase accountability [4]. This section summarizes the related state-of-the-art algorithms and approaches that have been taken to solve a gamut of problems in participatory budgeting.

Several methods have been proposed for ranking the project in PB. One such method is called the Technique for Order Preference by Similarity to Ideal Situation (TOPSIS) [5] attempts to diminish the gap between the situation under consideration and the ideal alternative while enlarging the gap between the situation under consideration and the worst situation. Further, fuzzy TOPSIS [6] which is an extension of the aforementioned work is proposed.

Reference [7] delves into the current models of participatory budgeting, aiming to align them with different sets of contextual factors and assesses the feasibility of implementing participatory budgeting in the emerging democracies of Central and Eastern Europe (CEE). Reference [8] dissected participatory budgeting into its communicative and empowerment components, highlighting that the empowerment aspect has often been overlooked in its worldwide adoption, which raises concerns regarding emancipation. The paper further explores the institutional reforms linked to empowerment in its initial concept, along with its analytical aspects.

Reference [9] demonstrates the effect of participatory budgeting on the political environment in Porto Alegre while also supporting the influential players in the field. This study also outlines accountability procedures that made it easier to elect council members with special resources, resulting in a situation where "dominated-dominants dominating the dominated" occurs. Most of the existing models that are currently in use for participatory budgeting take into account a fixed budget, however, the model presented in [10] also takes volatility into account. Reference [11] highlights the roots and characteristics of participatory budgeting and also discusses the future of Participatory Budgeting in the United States.

Since the project selection takes place through voting, multi-winner K scoring rules [12] such as Phragmen's Sequential rule [13] and Condorcet rules [14] are considered for budget allocation. There are various ways in which participatory budgeting can be seen for example as a social choice. The idea of handling social choice is discussed in [15] which becomes a suitable parameter in participatory budgeting.

Participatory budgeting can also be seen as a computational problem of choosing a subset of alternatives that maximizes utilitarian welfare in a constant finite budget. While studying the algorithmic properties of these methods, it has been observed that they are NP-Hard whenever the individual preferences are additive [16]. Even when the voter is awarded constant utility if and only if at least one of the approved alternatives is accepted in the final set of alternatives, thus, NP-hardness comes inherently [17]. The mathematical introduction of the concept of proportional representation [18] gives a landmark understanding of the idea and method for measuring the happiness of the voters in a particular allocation scheme.

Reference [19] highlights two logics namely management and community-building that have emerged throughout the development of participatory budgeting. These logics either coexist with or replace the conventional political logic. Therefore, within political institutions, these different logics may blend to varying degrees, creating a hybridized logic with varying degrees of harmony or conflict. Further, reference [20] provides a comprehensive survey of the various ideas explored by artificial intelligence scientists in the context of participatory budgeting.

After performing the literature review, we propose a new participatory budgeting scheme by using Sequential Monroe Rule (SMR) and Sequential Chamberlin Courant Rule (SCCR). The results obtained are significant and offer better allocation of budgets, moreover, the performance of rules in budget allocation is high and satisfies more than 90% of the criteria, thus, making the budgeting process more effective.

## 3. Methods and Materials

In this work, we propose to adapt two polynomial time computable rules, namely, Sequential Monroe Rule and Sequential Chamberlin Courant Rule to the settings of participatory budgeting. Particularly, we investigate their behavior in two cases 1) a special case where the prices of all the projects are valued at the same cost, and 2) projects may be of different costs. We verify the effectiveness of these rules by checking if their outputs satisfy the notion of proportional representation. In essence, we investigated the behavior of these preference aggregation rules based on the notion of Budget Justified Representation (BJR) (a PB variant of Justified Representation) in our scenarios. We present our results by experimentally verifying the results with different impartial cultures for both the aforementioned scenarios. We aim to conclusively state the better of the two well-known efficient polynomial time preference aggregation rules over their ability to select proportionally representative budgets in cases where the projects are either equally valued or diversely valued.

An important aspect of this study is the generation of voter preferences - we do this by generating votes of voters, candidate project sets, budgets, and corresponding costs of candidate projects uniformly at random. This is one of the most impartial ways of generating budgets and is rightly called *Impartial Culture*. We generate several instances of preferences and other important parameters such as candidate project sets, costs, and budgets, and set them as inputs to our algorithms. Table 1 presents the notation employed for the representation of this study and algorithms.

Table 1: Notation Utilized.

| Notation | Meaning |
| --- | --- |
| $\Phi_\alpha^S$ | Partial assignment that assigns a single alternate to at most $\left\lceil \frac{n}{K} \right\rceil$ agents |
| $l_1^\alpha(\Phi_\alpha^S)$ | Utilitarian satisfaction |
| $\Phi$ | Map defining the partial assignment |
| $\Phi^\leftarrow$ | Set of agents with defined assignment |
| $\Phi^\rightarrow$ | Set of agents with alternative assignment |
| $C$ | Candidate Set |
| $\succ$ | Stick Preference |
| $q$ | Quota |
| $L$ | Budget Limit |
| $c$ | Cost Function |
| $N$ | Set of Voters |
| $P$ | Set of Projects |
| $X$ | Budget (Set of Projects) |
| $A$ | Approval ballot of a voter |
| $R$ | Set of Real Numbers |
| $N'$ | Set of Natural Numbers |
| $JR$ | Justified Representation |

### 3.1 Preliminaries

Let, the set of voters be denoted by $V = \{V_1, V_2, …, V_n\}$ and the set of projects eligible to be funded be denoted by $P = \{P_1, …, P_m\}$. It is to be noted that we assume that the number of voters is $n$ and that the number of projects is $m$. We denote a budget by $X$, i.e. a subset of the projects $P$. Let, the voters' approval ballots be denoted by $A_i$, for all $i \in V$. Denote the cost function of a project by $c: P \rightarrow R$, where $R$ is the set of real numbers. Therefore, $c(P_i)$ represents the cost of the $i^{th}$ project. With little abuse of notation, we denote the cost of a set of projects $P'$ to be $c(P') = \Sigma_{i \in P'} P_i$. Further, we denote the budget limit by $L$, and define a budget to be *feasible* if the sum of the costs of all the projects in the budget sum up to at most $L$. The Fig. 1 shows the allocation of budget based on preferentially voting. Here, we have budget limit of $200, four projects (Bank, Park, Nursery, School), four voters to cast their votes. After, receiving voting statistics, a suitable algorithm is applied to select projects based on the available budget and voting preferences. Finally, two projects are selected i.e. School and Park.

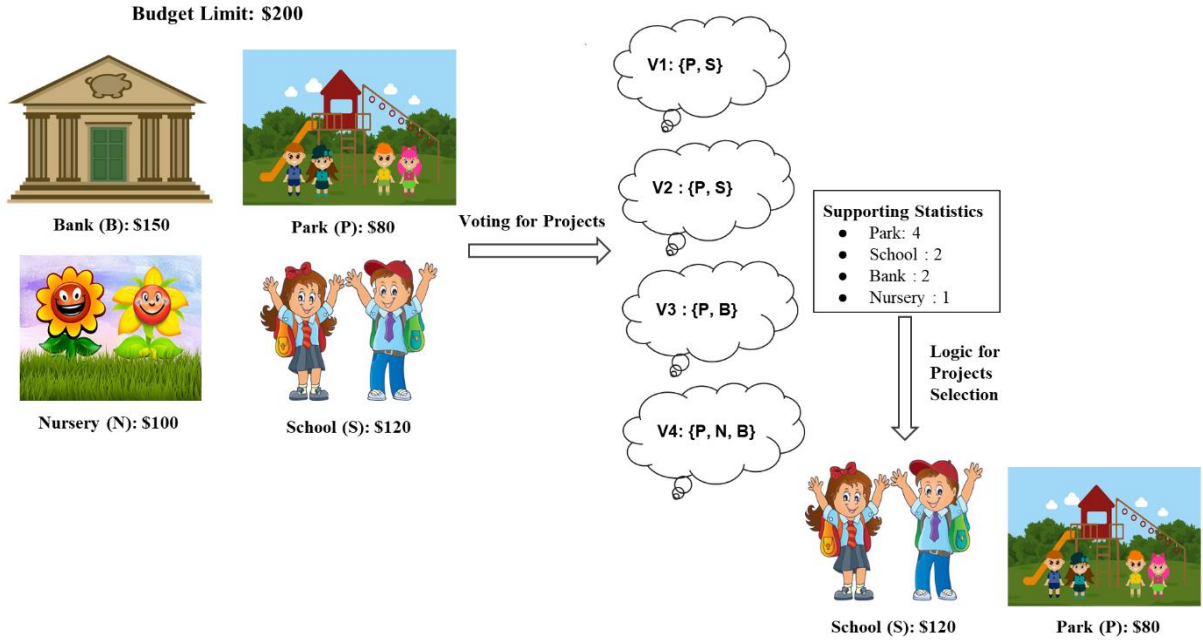

Figure 1: Example of budget allocation for different projects based on voter preferences.

### 3.2 Definitions

In this section, we define the proportionality notions of U-Justified Representation and Strong B-Justified Representation to lay down the basis for testing the nature of representation that the presented preference aggregation algorithms provide. In the case of U-Justified Representation, we assume that the project costs are uniform (and that there are $L$ projects) and equal to one unit universally; however, we make no such assumption in the case of Strong B-Justified Representation.

***Definition 1 (U-Justified Representation)*:** A budget $X$ satisfies *U-Justified Representation* if for every group of voters of size at least $n/L$ having at least one project in the intersection of their approval ballots, there is at least one project funded which is favored by at least one voter out of the voter group.

This definition essentially asserts the groups of voters that are of a deserving size and have fairly similarly aligned preferences must have one of their preferred projects funded. A fairly generalized definition for the case of non-uniform project prices can be found as follows (definition by Aziz et al. [13]).

***Definition 2 (Strong B-Justified Representation)*:** A budget $X$ satisfies *Strong B-Justified Representation* if for every voter group that agrees on funding at least one project and is of size at least $n/L$ has a non-zero cost project funded in the budget.

Note that Definition 2 is essentially *stronger* than Definition 1 in the sense that Definition 2 provides a representation to all those voter groups that get such in Definition 1. We define an *exhaustive budget* to be one in which no more projects can be filled without disregarding the given budget limit $L$.

***Definition 3 (Budgeting Method)*:** We define a *budgeting method* or a *preference aggregation* method as an algorithm that receives the tuple $<P, V, c, L>$ and outputs a budget allocation $X$, wherein, we expect $X$ to satisfy some proportionality notions.

### 3.3 Preference Aggregation Rules

***Equally Valued Projects*:** In most cases, the setting of "*Equally Valued Projects*" is a natural variant of the *Committee Selection Scenario* in the social choice settings. That is, the problem of selecting a subset of projects with a given budget limit from a set of equally valued projects is equivalent to electing a committee of size numerically equal to the budget limit. We inspect three well-known preference aggregation mechanisms namely, Sequential Monroe Rule, Sequential Chamberlin Courant Rule, and Single Transferrable Vote with inputs in the form of equally valued projects.

***Unequally Valued Projects*:** These projects form the generalized version of the equally valued projects. Such settings allow us to model the problem in a more realistic setting. Further, we run the Sequential Chamberlin Courant Rule variant of the algorithm for inputs in the form of unequally valued projects and tabulate the results.

In this paper, we only look at the case of participatory budgeting which has *Unequal Valuation* for Algorithm 2. Additionally, we look at *Equal Valuations* for all other Algorithms that we propose further. Essentially, all projects have an equal value i.e., unity. We adapt rules from Skowron et al. [21] and the algorithms used are

Algorithm 1 and Algorithm 2. Essentially, Algorithm 1 and Algorithm 2 represent the Sequential Chamberlin Courant Rule and Sequential Monroe Rule, respectively.

*Sequential Monroe Rule (Algorithm 1):* In this algorithm, we construct the budget iteratively i.e. we add a project to the budget and continue to do so until the inclusion of a project in the next iteration in the budget would render it unfeasible. Essentially, in SMR, we add a not-in-the-budget project $P_i$ to the budget if it lies in the approval sets of at least $n/k$ voters and lies in the approval sets of the maximum number of voters. Further, we eliminate these satisfied voters from consideration and re-iterate. We terminate the algorithm when the condition $c(P) \leq X$ turns false, where $P$ is the budget formed. Theoretically, this algorithm is a polynomial time $(1 - (X - 1) / 2(m - 1) - (H_k/k))$ approximation algorithm, where $H_k$ is the $k^{th}$ Harmonic Number.

---
**Algorithm 1: Sequential Chamberlin Courant**

$S \leftarrow \emptyset$
for $i \leftarrow 1$ to $K$ do
$\quad a \leftarrow argmax_{a \in A|S} l_1^\alpha \left( \Phi_\alpha^{S \cup \{a\}} \right)$
$\quad S \leftarrow S \cup \{a\}$
Return $\Phi_\alpha^S$

---

*Sequential Chamberlin-Courant Rule (Algorithm 2):* This algorithm, like **Algorithm 1,** is also iterative; however, in this case, we do not attach a lower limit of $n/X$ for the minimum number of voters to be satisfied. Rather, the algorithm starts by visiting every not-included-in-the-budget project and finds the number of voters that have the project in their approval sets. At the end of this step, the algorithm selects a project that has the maximum popularity, includes it in the budget, and eliminates satisfied voters from further consideration. The algorithm then checks if the set of selected projects doesn't overshoot the budget and if not, moves to the next iteration.

---
**Algorithm 2: Sequential Monroe Rule**

if $K \leq 2$ then
$\quad$ compute optimal solution use [22] and *return*
$\Phi = \{\}$
for i $\leftarrow$ 1 to K do
$\quad score \leftarrow \{\}$
$\quad bests \leftarrow \{\}$
$\quad$ foreach $a_i \in A \setminus \Phi^\rightarrow$ do
$\quad\quad agents \leftarrow sort\ N \setminus \Phi^\leftarrow$
$\quad\quad bests[a_i] \leftarrow chose\ 1st\ \left\lceil \frac{n}{K} \right\rceil\ elements\ from\ agents$
$\quad\quad score[a_i] \leftarrow \sum_{j \in bests[a_i]} \left( m - pos_j(a_i) \right)$
$\quad a_{bests} \leftarrow argmax_{a \in A \setminus \Phi^\rightarrow} score[a]$
$\quad$ foreach $j \in bests[a_{best}]$ do
$\quad\quad \Phi[j] \leftarrow a_{best}$

---

*Standard Transferrable Vote (STV):* STV is a widely utilized algorithm for preference aggregation since it satisfies many strong variants of Proportional Representation. However, it is a non-monotonous rule i.e. if some candidate $a$ is preferred over $b$ by all voters, STV might elect $b$; such rules are called non-monotonous rules. STV also is an iterative algorithm and it elects budgets by iteratively adding those projects to the budget that have the highest approval and then eliminating them from the approval sets of the satisfied voters. The algorithm for STV is shown in Algorithm 3.

---
**Algorithm 3: Single Transferable Vote ($N, C, k, q, \succ$)**

$W \leftarrow \emptyset$
$w_i \leftarrow 1$ for each $i \in N$
$j \leftarrow 1$
while $|W| < k$ do
$\quad$ if $|W| + |C(\succ)| = k$ then
$\quad\quad$ return $W \cup C(\succ)$
$\quad$ if candidate $c$ with plurality support at least $q$ then
$\quad\quad$ Voters supporting c will be $N'$. Update weights of voters in $N'$, thus, total weight in $N'$ lessen by $q$.
$\quad\quad$ Delete $c$ from $\succ$
$\quad\quad W \leftarrow W \cup \{c\}$
$\quad$ else
$\quad\quad$ Delete candidate with least plurality from current preference $\succ$
*return W*

## 4. Results and Discussions

We discuss the results of our experiments in Table 2, where we report the probability of different algorithms to satisfy the U-Justified Representation axiom. We divide the results into two parts - 1) projects are equally valued, 2) and projects are unequally valued with no assumption. For the generation of preferences, we use the Impartial Culture Method of preference generation in which we generate the preferences of the voters randomly while keeping the costs of projects and the corresponding budget limits coherent, i.e., not allowing the budget limits to be lesser than the costs of individual projects.

The experiment is performed on Google Colab, with available RAM of 12 GB and the programming language is Python. Specifically, with n1-highmem-2 instance 2v and CPU@2.2 GHz with a free space equivalent to 64 GB.

In the case of *equally valued projects*, we deploy three preference aggregation rules mentioned in Table 2, i.e., Sequential Chamberlin Courant Rule, Sequential Monroe Rule, and Quota-Single Transferrable Vote Rule. The general procedure followed is the generation of different numbers of participatory budgeting instances and finding corresponding budgets with respect to the given preference aggregation rules. After this, we check whether these outputs/budgets satisfy the U-Justified Representation axiom for each of the instances generated and find the probability with which the given algorithms satisfy the axiom. We present a comparison of all these probabilities in Table 2.

The results obtained have been registered in Table 2. We performed experimentation by generating different numbers of election instances, given different conditions where objects are either equally priced or unequally priced. We run the polynomial variants of Chamberlin Courant, Monroe, and Standard Transferrable Voting rules for the equally spaced version and find the probability that the generated budget follows the proportional representation axiom. Moreover, we run the polynomial version of Chamberlin Courant for the *general case (Unequal Valuations)* where the projects to be included in the budget are differently priced.

Table 2: Probabilities of different algorithms satisfying axiom of U-Justified Representation.

| No. of Instances | Equally Valued | | | General Case |
|---|---|---|---|---|
| | SCCR | SMR | STV | SCCR |
| 100 | 94.39 | 93.98 | 95.56 | 67.90 |
| 300 | 93.83 | 92.21 | 94.78 | 56.19 |
| 500 | 94.21 | 92.09 | 96.69 | 54.10 |
| 1000 | 92.22 | 90.19 | 95.77 | 49.66 |
| 3000 | 94.39 | 89.32 | 96.49 | 48.12 |
| 5000 | 94.10 | 90.12 | 94.99 | 50.21 |

From Table 2, it is evident that all the proposed budgeting methods perform quite well when the objects that are put to budgeting are of the same price. That is, the probabilities are all higher than 90%, as compared to the case where the objects are unequally priced, where the probabilities are less than even 70%.

From Fig. 2, is also to be noted that, the general trend of satisfaction of the axiom is decreasing when measured in the order of increasing number of sequences. This may primarily be so because as the number of instances generated increases, the number of different scenarios generated also increases and therefore, the U-Justified Representation axiom becomes more difficult to satisfy.

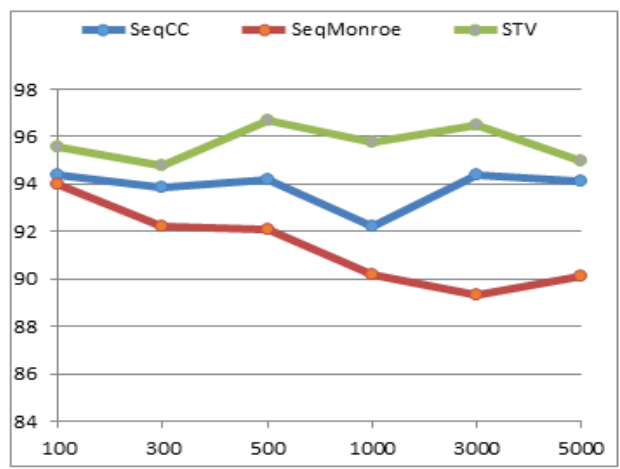

Figure 2: Probabilities of axiom satisfaction for equal-valued projects

Moreover, we observe that the highest probability of satisfaction of the U-Justified Representation axiom (in the *special case i.e. Equal Valuations*) is with the voting rule Standard Transferrable Voting (STV), followed by the Sequential Version of Chamberlin Courant Rule, followed by Sequential Monroe A rule. This is in coherence with the generally accepted notion of STV being one of the most suitable voting rules for proportional representation, though, it misses out on the satisfaction of some very crucial properties like monotonicity. We also observe the fact that the Sequential Monroe A rule performs the worst among all the other rules that we implement in this regard. This is primarily because the Chamberlin Courant Rules and Monroe Rules are families of voting rules that are fundamentally used for different purposes, one of them having to do with the election of a 'diverse committee'.

Furthermore, we implement only the Sequential Chamberlin Courant rule for the general case, whose results have been documented in Table 2 and probability has been plotted in Fig. 3. We find that its performance in comparison with the case when all the costs of objects are unequal is quite poor, ranging around 68%. The general case therefore is much more difficult to realize proportional representation and therefore more difficult to deal with.

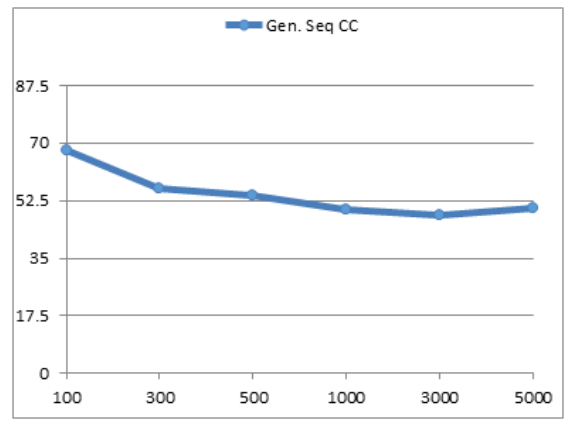

Figure 3: Generalised Sequential Chamberlin Courant Rule axiom satisfaction.

## 5. Conclusion and Future Work

We conclude from the discussion about the general unreliability of good limits of proportional representation, even with the use of state-of-the-art voting rules, more so, when the costs of the objects to be funded are equal. In effect, we compared different voting rules, which are well known for producing proportionally representative budgets in two different conditions, i.e. the case where the projects that have to be moulded into budgets are equally valued and the case where they are differently valued, not necessarily drawn from a special distribution. We show that as the number of situations that a voting rule faces increases, the probability that it satisfies a proportionally representative axiom decreases, empirically proving that as the number of situations represented by randomly generated voter profiles increases, the voting rules become more prone not to satisfy the presented proportional representation axiom. However, we do perform a comparative analysis of the performance of the voting rules, which we do essentially by comparing the probability with which they satisfy the proportionality axiom for different numbers of voter profiles.

We conclude that the best-performing rule for the case of equally valued projects is Standard Transferrable Vote, followed by Sequential Chamberlin Courant, followed by Sequential Monroe Rule. The performance of the rules in general is quite high, ranging to greater than 90%, therefore making the budgeting process effective. Moreover, we also conclude that the Sequential Chamberlin Courant rule for the case of differently valued projects is quite non-performing, with results ranging to 50.21% in the worst case. From this, we conclude that the budgeting process is quite difficult and inefficient when the preference aggregator used is the Sequential Chamberlin Courant rule.

We also infer that checking for the proportionality of the budget becomes increasingly complex when the number of randomly generated instances increases. It is already known that checking the proportionality of a budget is Co-NP hard and therefore, as the number of instances input increases, the time required to run the algorithm increases. In this way, we also present computational limitations of checking the proportionality of the budget, which is also an important contribution.